\documentstyle[11pt,newpasp,twoside,epsf]{article}
\def\edcomment#1{\iffalse\marginpar{\raggedright\sl#1\/}\else\relax\fi}
\reversemarginpar

\begin{document}
\title{Properties of the Non-Thermal Emission in Plerions}
\author{Rino Bandiera}
\affil{
Osservatorio Astrofisico di Arcetri,
Largo Fermi 5, 50125 Firenze (Italy)}

\begin{abstract}
The synchrotron emission observed in plerions depends on the characteristics of
the magnetic fields and relativistic electrons in these supernova remnants.
Therefore an analysis of the spectral and spatial properties of this emission,
combined with models for the evolution and structure of plerions, would allow
one to investigate the evolution of the synchrotron nebula, the structure of
the magnetic field, the distribution of the relativistic electrons, as well as
to put contraints on the history of the energy input from the associated
neutron star.

The identification of a new class of plerionic remnants, with spectral
properties different from the Crab Nebula, has been proposed. The spectra of
these objects typically show a sharp spectral break at very low frequencies
(below 50 GHz), with a steep spectrum right beyond the break. In order to model
the properties of their emission, a non-standard evolution of the pulsar output
seems to be required.

X-ray observations of the synchrotron emission from the Crab Nebula are shown.
They are compared with previous data, and their implications on the structure
of the Crab are discussed. Recent millimetric data of this object are also
presented. A spatially resolved analysis, based on radio, millimetric and X-ray
data, will be carried on also for other plerions.
\end{abstract}


\section{Introduction}

Not all supernova remnants (SNRs) show the ``classical'' shell-like structure.
Some of them present instead a filled-center structure: the prototype is the
Crab Nebula, but other objects with similar properties have been discovered
since then. They are called filled-center SNRs, or Crab-like SNRs, or {\it
plerions} (Weiler \& Panagia 1978).
Plerions are recognized not just on the basis of their morphology, but also
of other properties, like:
{\it a flat power-law radio spectrum,} with a spectral index ranging from 0.0
to $-$0.3;
{\it a high radio polarization,} with a well organized pattern (not true for
all plerions);
{\it a power-law X-ray spectrum,} with photon index close to $-$2
(Asaoka \& Koyama 1990);
{\it the detection of an associated pulsar} (not true for all plerions --- see
Pacini, these Proceedings).

Although the details of the nature and structure of plerions are still unclear,
there is a common agreement on the following points:
{\it a plerion is an expanding bubble, formed essentially by magnetic fields
and relativistic electrons,} and the observed synchrotron emission originates
from these two components;
{\it a continuous supply of magnetic flux and relativistic particles is
required,} in order to explain the typical synchrotron emissivities, as well as
the high frequency emission (from particles with synchrotron lifetimes shorter
than the SNR age).

In a simplified approach, one may assume that magnetic fields and particles are
uniformly distributed into the plerionic bubble: this approach (Pacini \&
Salvati 1973; Reynolds \& Chevalier 1984; Bandiera et al.\ 1984) is usually
adequate to explain the evolution of the overall nebular spectrum. However the
homogeneity assumption is likely to be incorrect. New particles and magnetic
flux are released by the associated pulsar, presumably near the pulsar wind
termination shock (Rees \& Gunn 1974; Kennel \& Coroniti 1984a,b).
Therefore the degree of homogeneity of the electrons distribution depends on
how efficient are the mechanisms (diffusion, or advection) by which particles
propagate through the nebula (see Amato, these Proceedings).

Also the structure of the magnetic field can be rather complex. From
considerations on the MHD relations it follows that spherical models cannot
account adequately for the field structure: it can have at most a cylindrical
symmetry (Begelman \& Li 1992), but more complicate patterns are suggested by
observations. The comparison of high resolution maps at various frequencies may
then give important clues on the structure on plerions, and on the processes
governing the evolution of the magnetic field as well as of the particle
distribution.

The outline of this paper is the following: I begin by reviewing the
classical, simplified approach to the evolution of a plerion and of its
emission; then I consider the case of the Crab Nebula, the prototype of this
class, as well as of some other plerions with characteristics different from
the Crab; finally I describe the results and perspectives of a multifrequency
study, with high spatial resolution, of the Crab Nebula and ot and other plerionsher plerions.


\section{Classical models of the evolution of the synchrotron emission}

Classical models for the evolution of the synchrotron emission from a plerion
as a whole (Pacini \& Salvati 1973; Reynolds \& Chevalier 1984; Bandiera et
al.\ 1984) are based on the original analysis of Kardashev (1962). The starting
points are the two basic synchrotron equations, that for the radiated power
from an electron with energy $E$ ($W_s=c_1B^2E^2$), and that for the typical
frequency of radiation ($\nu_s=c_2BE^2$): these formulae are averages over
pitch angles, under the assumption of isotropy. If $N(E)$ is the present
distribution of electrons, and the magnetic field $B$ is constant throughout
the nebula, it immediately follows that the synchrotron spectrum is given by
$L(\nu)=(c_1/2c_2)BEN(E)$. This relation establishes the well known connection
between the particle distribution and the radiated spectrum (if
$L(\nu)\propto\nu^{-\alpha}$, then $N(E)\propto E^{-(1+2\alpha)}$).

Let us consider, for instance, the synchrotron spectrum of the Crab Nebula.
One may identify various spectral regions with different spectral indices: the
radio, with $\alpha\simeq0.3$; the optical, with $\alpha\simeq0.8$; the X rays,
with $\alpha\simeq1.0$; and a further steepening above 100 keV. All the changes
of slope in $L(\nu)$ correspond to breaks in $N(E)$: the issue is to determine
which of them are intrinsic to the injected particle distribution, and which
result from the evolution. Usually an original power-law distribution is
assumed, with the aim of explaining all breaks as originated just from the
evolution.

The basic ingredients are:
1. {\it The evolution of the energy input from the spinning down pulsar} (this
input typically lasts for a time $\tau_o$, and then falls down).
2. {\it The fraction by which this power is shared between injected particles
and field} (usually assumed to be constant).
3. {\it The expansion law of the nebula,} $R(t)$, that at earlier times may
be linear or even accelerated; but later on, with the passage of the reverse
shock coming from the outer blast wave, the plerion may shrink, and then
re-expand at a lower rate (Reynolds \& Chevalier 1984).

The evolution of the magnetic field in the nebula is modelled by including the
effects of the adiabatic losses; while for the evolution of particles both
adiabatic and synchrotron losses must be taken into account.
A special particle energy is that at which the timescales for adiabatic and
synchrotron losses are comparable ($E_b\sim1/c_1B^2t$): a break in the
spectrum occurs at the frequency that corresponds to $E_b$.
Before the time $\tau_o$ this is the only evolutive
break present in the spectrum. Kardashev (1962) showed that the change in the
spectral index must be $\Delta\alpha=0.5$: this result can be directly tested
on the data.
After the time $\tau_o$, a second break should appear in the distribution, at
the energy $E_c(t)=E_b(\tau_o)R(\tau_o)/R(t)$: this is the ``fossil break'',
namely the adiabatic evolution of the break located at $E_b$ at the time
$\tau_o$. This break is at a frequency smaller than that of the truly evolutive
break.

The Crab Nebula fits rather well into the scenario described above. In fact:
1. {\it The classical law for the pulsar spin-down seems to be verified:} the
total energy released by the pulsar since its birth ($\sim10^{49}$~erg) is also
consistent with the kinetic energy excess in the optical thermal filaments (due
to their dynamical coupling with the plerionic bubble).
2. {\it The efficiency in particle production is reasonably high:} the present
pulsar spin-down power ($4.5\times10^{38}\,{\rm erg\,s^{-1}}$) is comparable
with the total synchrotron luminosity ($\sim0.7\times10^{38}\,{\rm
erg\,s^{-1}}$).
3. {\it The efficiency in magnetic field production is rather high:} from the
position of the break ($\nu_b\simeq10^{13}$~Hz) one may derive a nebular
magnetic field $B\sim0.4$~mG, i.e.\ a magnetic energy $\sim2\times10^{49}$~erg,
close to the total energy released by the pulsar.
4. {\it The measured secular variation in radio} ($-$0.17\% per yr; Aller \&
Reynolds 1985) {\it agrees with the theoretical estimate} (V\'eron-Cetty \&
Woltjer 1991).
5. {\it The change in the spectral slope from radio to optical is
$\Delta\alpha=0.5$,} in agreement with Kardashev; however the further breaks at
higher frequencies cannot be explained in this way, unless breaks in the
injected distribution are invoked.

It can be noticed that in the Crab Nebula fields and electrons are in near
equipartition. Is this equipartition typical for all plerions? Does
equipartition hold at injection, or is it the result of a subsequent
field-particle coupling?


\section{Non Crab-like plerions}

There is a bunch of plerions characterized by a spectral break at frequencies
much lower than in the Crab Nebula: for instance, in 3C58 and in G21.5$-$0.9
the break is at 50~GHz; while in CTB87 it is at 20~GHz. Woltjer et
al.\ (1997) discussed properties and evolutive implications for these objects.
Beyond the low-frequency position of the break, they share other features, like
a sharp break, with $\Delta\alpha$ larger than the canonical value 0.5.
Furthermore, in none of these objects pulsations from the expected neutron star
have been detected yet.

It is hard to explain the observed break as the main evolutive break, since it
would imply a very large nebular field. The most extreme case is that of CTB87,
an extended plerion for which the estimated magnetic energy is greater than
$6\times10^{51}$~erg, well above that of the supernova explosion itself. This
paradox can be solved if the break observed is the fossil break: in this case
also the theoretical limit on $\Delta\alpha$ across the break can be overcome,
under the condition that the pulsar slow-down follows a very steep law (even
though some difficulties remain, in modelling the break sharpness). This
implies that the associated pulsar has slowed down considerably,
and therefore has become much fainter than at the origin: this
may be a reason why no pulsations are detectable.

Among these plerions, 3C58 is that posing more problems to models. In this
object the sharpness of the break requires an abrupt decrease in the rate of
injected particles. Beyond that, models must also account for the measured
increase in the radio emission (Green 1987, and references therein). Woltjer et
al.\ (1997) show that, in order to match the observations, a sudden change in
the relative efficiencies of field and particles production is required: this
could actually be associated to a recent ``phase change'' in the pulsar
magnetosphere.


\section{Plerionic components in composite SNRs}

Composite SNRs are those in which a shell-like component (due to the
interaction of the supernova ejecta with the ambient medium) co-exists with a
plerionic component. Helfand \& Becker (1987) introduced the class of
composite SNRs and outlined their properties: they are old enough to have
developed a shell component, but still young enough to host a detectable
plerionic component.

Studying a composite SNR is more interesting than just studying independently a
shell-like and a plerionic remnant, since the two components have the same
origin (thus the same age and distance), and are interacting. Slane et
al.\ (1998) discuss what kind of diagnostic tools can be used. For instance,
the X-ray spectrum of the thermal shell allows one to evaluate its pressure and
therefore, by assuming a (rough) pressure equilibrium with the plerion, the
plerionic magnetic field. This is an estimate alternative to that based on the
position of the evolutive break, and may then represent a test on the nature of
that break.

In the cases in which the associated pulsar has not been detected yet, its most
likely parameters may be estimated. If a pulsar has been already detected, its
spin-down time may be used to estimate the SNR age, that should agree with the
age derived from the thermal X-ray spectrum. Moreover one can derive the
present pulsar energy output, and then the efficiency by which this
energy is converted into magnetic fields and particles.
This analysis has been already carried on for various composite SNRs,
like G11.2$-$0.3 (Bandiera et al.\ 1996), CTA1 (Slane et al.\ 1997), N157B
(Cusumano et al.\ 1998), MSH11$-$62 (Harrus et al.\ 1998), G327.1$-$1.1 (Sun et
al.\ 1999), G39.2$-$0.3 (Harrus \& Slane 1999).


\section{The Crab Nebula: results of a multifrequency analysis}

The above considerations on the Crab Nebula, as well as those on other
plerions, are mainly referring to their global properties. But a much deeper
insight should follow from a combined study of the spatial-spectral properties
of the synchrotron emission from these objects. I present here some results
coming from a comparison of optical and X-ray maps of the Crab Nebula, and a
preliminary analysis of millimetric observations of this plerion.

The Crab Nebula is a very bright object, over a wide range of frequencies:
therefore it is an ideal target for a detailed multifrequency investigation
with high spatial resolution. Our project has been inspired by the study of
V\'eron-Cetty \& Woltjer (1993) on the Crab synchrotron emission in the
optical range: they produced a map of the optical spectral index, and showed
that a spectral steepening occurs going outwards; they also noticed that
the region with optical spectral index flatter than $-$0.7 matches the
shape of the Crab in X rays.

We decided to carry on a detailed, quantitative comparison between optical and
X rays: for this reason we have re-analyzed the V\'eron-Cetty \& Woltjer (1993)
optical data, while for the X rays we have used public data of ROSAT HRI. In
the X-ray data reduction we have stressed, rather than the resolution, the
sensitivity to regions with very low surface brightness. For this purpose we
have deconvolved the map in order to eliminate the wings of the instrumental
Point Spread Function, as well as the halo due to dust scattering. Details on
the data analysis and on their interpretation are given by Bandiera et
al.\ (1998).

As already pointed out by Hester et al.\ (1995) the outer and faint X-ray
emission extends almost to the boundary of the optical nebula. We have then
performed a quantitative analysis over a large area, by using $5"\times5"$
pixels. Fig.~1 gives a plot of the optical-to-X averaged spectral index
($\alpha_{OX}$) versus the optical spectral index ($\alpha_{Opt}$), where each
dot refers to a single pixel: unfortunately no high resolution spectral map in
the X rays is available yet.

In the interpretation of this plot our underlining assumption is that, due to
the evolution, the spectral indices of the particle distribution, in all
energy ranges, tend to steepen: this corresponds to a softening of the emission
spectrum at all frequencies, and translates into the requirement that the time
evolution of a given bunch of particles produces a drift of the related dot
towards the upper right direction of the plot.

\begin{figure}
\plotone{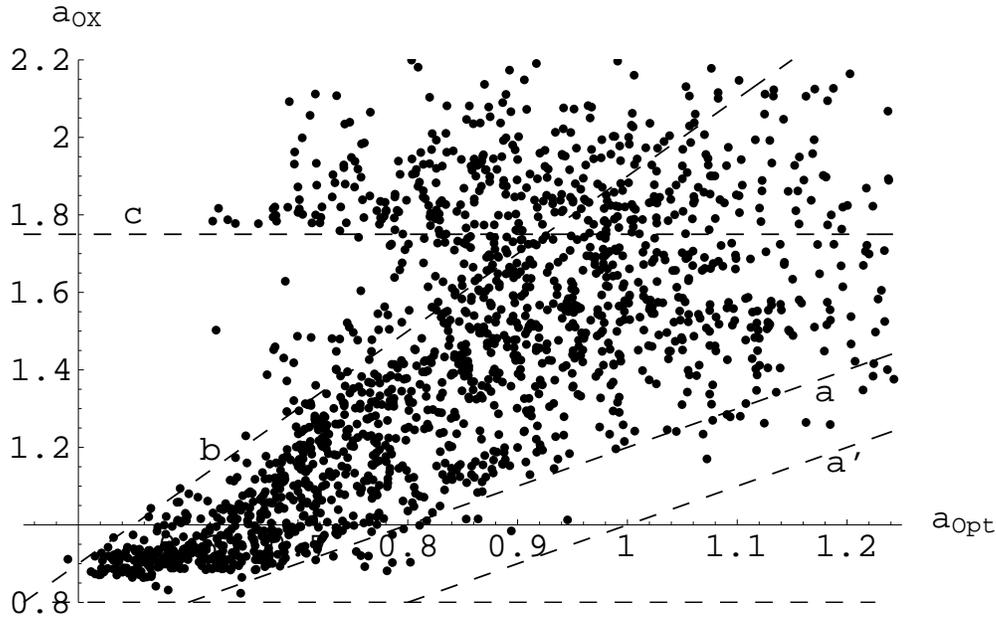}
\caption{
Plot of the $\alpha_{OX}$ versus $\alpha_{Opt}$ spectral index.
Each dot corresponds to a $5''\times5''$ pixel in the image of the Nebula.
See text for the meaning of the various lines.
}
\end{figure}

The area of the plot can be then subdivided into zones. Most of the dots are in
the region confined by the two lines labelled by $a$ and $b$: let us call it
the ``Main Region'' of the plot. All the pixels corresponding to the dots in
the Main Region are located in the main body of the Crab Nebula, namely that
obtained by cutting out the N-W and S-E elongations. The line
$a'$, parallel to $a$, indicates the cases of no spectral bending between
optical and X rays ($\alpha_{Opt}=\alpha_{OX}$). Pixels with $\alpha_{OX}<1$
outline rather well the region of the X-ray ``torus'' (see e.g.\ Hester at al.\
1995): in the plot they are located on the lower left side, in agreement with
our expectation that this is the main location where to find freshly injected
particles.

We introduced the quantity $m$, that
gives the position of a dot across the strip bounded by lines $a$ and $b$:
this quantity is related to the spectral bending between optical and X rays. A
map of $m$ is given in Fig.~2-L: points with lower $m$ (namely points with a
lower bending; brighter pixels in the map) are generally located in a thick
``equatorial'' belt. In particular, pixels with a very low bending coincide
with some prominent thermal filaments. If the absence of bending is a sign of
freshly injected, or re-accelerated, particles, then in these regions some
secondary acceleration processes could take place.

Let us go back to the plot of Fig.~1. Following the scheme introduced above,
the dots above line $b$ (``Secondary Region'') cannot be the result of a
continuous evolution from dots originally located in the Main Region.
They are more likely to represent the evolution of particles originally
emitting a spectrum with $\alpha_{OX}$ not smaller that 1.75
(line $c$), and that afterwards, while moving outwards, have softened
considerably
their spectrum. Fig.~2-R gives a map of $\alpha_{OX}$ for all the points
confined to the two lobes (brighter pixels indicate higher values of
$\alpha_{OX}$): here a softening of the spectrum in the outer zones is
apparent. In the lobes, the zones with harder spectra seem to lie on the
prolongation of the X-ray ``jets'' (see Hester et al.\ 1995). This is
very clear for the S-E lobe, and may be the indication that these
particles are directly provided by the jets.

\begin{figure}
\plotone{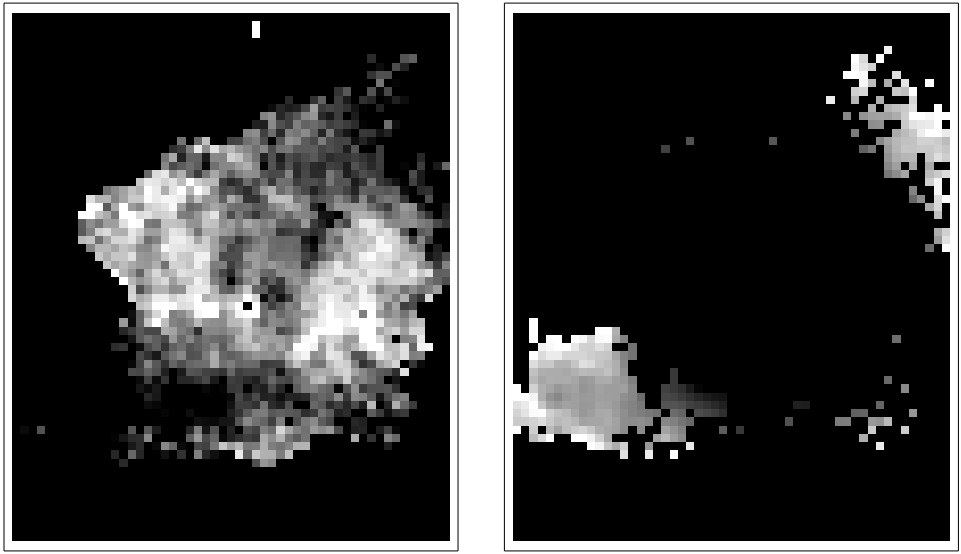}
\caption{
{\bf Left:} map of the quantity $m$, for the Main Region (regions with less
bent spectra are brighter).
{\bf Right:} map of $\alpha_{OX}$, for the Secondary Region (steeper spectra
are brighter).
}
\end{figure}

All these considerations are largely empirical. Moreover, some conclusions may
be partly biased by projection effects: anyway the general results
should remain valid. The first result is the great variety of
local spectra in the nebula. This may be partly explained in terms of the
evolution of electrons injected at the central torus; however, both in the
regions with low optical-to-X spectral bending and in the polar lobes further
particle components seem to be present, suggesting the
presence of secondary acceleration (or re-acceleration) processes.

A similar result follows from recent observations at 230~GHz, with the
IRAM 30~m telescope, in collaboration with
R.Cesaroni and R.Neri: a comparison with
a map at 1.4~GHz shows variations in the spectral index between the two
frequencies. If confirmed, this result is very important. In fact the
spectral index of the Crab Nebula in the radio shows very little spatial
variations (Bietenholz et al.\ 1997) and, if there is only one channel of
injection, the same behaviour is expected up to the
evolutive spectral break ($\sim10^{13}$~Hz for the Crab).
On the contrary, flatter spectra are found, between 1.4 and 230~GHz, in the
central regions of the Crab Nebula, similarly as seen
in the optical range. Further observations are however required, in order to
confirm this result.


\section{Conclusions}

The comparison of high-resolution maps at different frequencies may provide a
wealth of information on the plerions, and will help to clarify some open
issues on these objects, like: how, where and by how many different mechanisms
particles can be accelerated inside plerions? How do they propagate through the
nebula? What is the structure of the magnetic field? How the evolution of a
plerion and of the associated neutron star affect the synchrotron emission?

A big step towards
the understanding of these objects will hopefully come with the
arrival of the new generation X-ray telescopes, that will be able to perform
spectral mapping with arcsec resolution. X-ray emitting electrons have very
short lifetimes, typically of the order of the light crossing time of the
nebula, and therefore from an X-ray spectral mapping one could derive
information directly on the sites of the present-time injection.
But another promising spectral range for effectively probing the physical
conditions in the nebula is the millimetric one. In various plerions a
spectral break is located near that range: spectral maps may then give
indications on the spatial variations of the break position, and therefore
on the magnetic structure of the nebula.

Even in the presence of such high quality observations, one may wonder how
powerful the synchrotron emission is as a diagnostic tool. In fact it can
provide just a mixed information on particles and magnetic field, and only
projected along the line of sight. Therefore even the interpretation of high
resolution spectral maps is generally not straightforward, unless theoretical
models will be developed at a level of detail similar to that of the
observations.


\acknowledgements

Many of the results presented here come from discussions and collaborations
with various persons: E.Amato, F.Bocchino, R.Cesaroni, R.Neri, F.Pacini,
M.Salvati, P.Slane, L.Woltjer.
This work is partly supported by the Italian Space Agency (ASI) through grant
ARS-98-116.


\end{document}